\documentstyle[sprocl,amsmath,amssymb,graphicx]{article}
\newcommand{\GeV}{{\mathrm{GeV}}} 
\newcommand{\TeV}{{\mathrm{TeV}}}
\newcommand{\fb}{{\mathrm{fb}}}
\newcommand{\ii}{{\mathrm{i}}}
\begin{document}
\rightline{\vbox{\halign{&#\hfil\cr
TTP-04-11\cr
WUE-ITP-2004-015\cr
August 2004\cr}}}
\title{%
  TESTS OF THE NONCOMMUTATIVE STANDARD MODEL AT A FUTURE PHOTON
  COLLIDER\,\footnote{%
    Work supported by Deutsche Forschungsgemeinschaft (RU\,311/1-1),
    and Bundesministerium f\"ur Bildung und Forschung, Germany (05HT1RDA/6).}}

\author{%
  THORSTEN OHL\,${}^1$%
    \,\footnote{e-mail:~ohl@physik.uni-wuerzburg.de}
  and
  J\"URGEN REUTER\,$^2$%
    \,\footnote{e-mail:~reuter@particle.uni-karlsruhe.de}%
    \,\footnote{Presented~at~LCWS2004, Paris, France, April~2004}}

\address{%
  ${}^1$%
    Institut~f\"ur~Theoretische~Physik~und~Astrophysik,\\
    Universit\"at~W\"urzburg, D--97074~W\"urzburg, Germany\\
  ${}^2$%
    Institut f\"ur Theoretische~Teilchenphysik,\\
    Universit\"at~Karlsruhe, D--76128~Karlsruhe, Germany}

\maketitle
\abstracts{%
  Extensions of the Standard Model of elementary particle physics to
  noncommutative geometries are proposed by string
  models. Independent of this motivation, one may consider such a model
  as an effective field theory with higher-dimensional operators
  containing an antisymmetric rank-2 background field. We study the
  signals of such a Noncommutative Standard Model~(NCSM) and analyze
  the discovery potential of fermion pair production at a future
  photon collider.}


The interest in noncommutative~(NC) theories 
has been growing dramatically in recent years due to the observation that
open string theories with constant antisymmetric rank-2 tensor
fields can be interpreted as Yang-Mills theories living on a NC
manifold~\cite{SeiWit}. Independent from this
motivation, NC Quantum Field Theory in itself provides an interesting  
approach in introducing a fundamental length scale and 
cutting off short-distance contributions, consistent with
the symmetries of a given model. Recently, there has been a lot of
effort in constructing an Effective Field Theory~(EFT) which is
defined on a NC spacetime with a canonical structure  
\begin{equation}
\label{eq:theta}
  [\hat{x}^\mu, \hat{x}^\nu]
    = \ii \theta^{\mu\nu}
    = \ii \frac{1}{\Lambda_{\text{NC}}^2} C^{\mu\nu}
    = \ii \frac{1}{\Lambda_{\text{NC}}^2} 
    	 \begin{scriptsize}
    	   \begin{pmatrix}
    	       0 & - E^1 & - E^2 & - E^3 \\ 
    	     E^1 & 0     & - B^3 &   B^2 \\
    	     E^2 &   B^3 & 0     & - B^1 \\
    	     E^3 & - B^2 &   B^1 & 0
    	   \end{pmatrix}
    	 \end{scriptsize}
\end{equation}
and which has the Standard Model~(SM) as low energy limit
for~$\sqrt{s}\ll\Lambda_{\text{NC}}$. The noncommutativity
$\theta^{\mu\nu}$ is an antisymmetric real matrix, assumed here to be
constant and understood as a spurion, breaking Lorentz invariance. Its
origin is irrelevant as long as we are merely studying
the EFT, where it appears as a coefficient in front of operators of
dimension six or higher. The crucial theoretical point
is the realization of gauge invariance on NC spaces
\cite{SeiWit,Wess:pr}. A general way is provided by the Seiberg-Witten
Map~(SWM)~\cite{SeiWit}. It is an asymptotic expansion in the
noncommutativity~$\theta^{\mu\nu}$, which expresses the NC matter
and gauge fields as functions of commutative matter and gauge
fields so that the NC gauge transformations are realized by the
commutative ones.  

In the Noncommutative Standard Model (NCSM)~\cite{NCSM}, one has
to replace each SM field by the corresponding SWM and all products by
$\star$-products (which realizes products of noncommutative functions
in terms of commutative ones). There are
higher-dimensional operators, whose contributions are suppressed by
the ratios $\Lambda^2_{EW}/\Lambda^2_{\text{NC}}$ or
$s/\Lambda^2_{\text{NC}}$ of the EW scale or the CMS energy
and the NC scale, giving rise to deviations of SM decay rates and  
production cross sections. SM vertices receive
corrections with new Lorentz structures, and there will be contact
terms of the form $f \bar f \gamma\gamma$ etc., which are required by
gauge invariance. In addition there can also be triple neutral
gauge boson couplings (TGC) like $\gamma\gamma\gamma$, $Z\gamma\gamma$, 
etc. \cite{TGC}. Current bounds (mainly from the non-observation of
the Lorentz-violating $Z$ decays $Z\to \gamma\gamma, gg$ at LEP) are
$\Lambda_{\text{NC}} \gtrsim 100\; \GeV$. In a low-lying string
scenario one could expect values for such a scale as low as
$\Lambda_{\text{NC}} \gtrsim 1 \,\TeV$.   

In the NCSM, amplitudes are asymptotic expansions in
$\theta^{\mu\nu}$. In lowest order in $\theta$, for
processes forbidden in the SM, the square of the first order NCSM amplitude
is the leading contribution, while for a nonvanishing SM amplitude
the deviation of the leading order is given by the interference of the
SM with the lowest order NCSM amplitude. Higher orders require
the (yet unknown) second and higher orders in the SWM. 

Our proposal is to study fermion pair production $\gamma\gamma \to
f\bar f$ at a future photon collider. Such a machine is planned to be
operated at a future linear $e^+ e^-$ collider with energies up to $1
\,\TeV$. Highly energetic photons are produced by Compton
backscattering of laser photons off the electron beam.
The photons can be delivered with a high degree of polarization, which 
is crucial for our considerations.

Theoretically, producing massless fermions ($\sqrt{s} \gg m_f$)
from polarized photons suggests to use the very concise
formalism of helicity amplitudes for evaluating the cross section.
The noncommutativity $\theta^{\mu\nu}$ can elegantly be converted to
a spinor  expression containing only the spinor metric $\epsilon_{AB}$
and a symmetric rank-2 spinor~$\phi_{AB}$
\begin{equation}
  \theta_{A\dot{A},B\dot B}
     = \theta^{\mu\nu}
        \bar{\sigma}_{\mu,A\dot A}
        \bar{\sigma}_{\nu,B\dot B} 
     = \phi_{AB} \epsilon_{\dot A\dot B}
        + \bar{\phi}_{\dot A\dot B} \epsilon_{AB}
\end{equation}
with $\phi_{AB} = (\bar\phi_{\dot A\dot B})^*$ and three
independent complex components
\begin{equation}
  \phi_{11} = - E_- - \ii B_-,\;
  \phi_{12} = E_3 + \ii B_3 = \phi_{21},\;
  \phi_{22} = E_+ + \ii B_+\,,
\end{equation}
where $E_{\pm} = E_1\pm \ii E_2$ and $B_{\pm} = B_1\pm \ii B_2$.
Using this decomposition, all amplitudes can be expressed as contractions
of Weyl-van der Waerden spinors with~$\phi$ and among themselves.

\begin{figure}
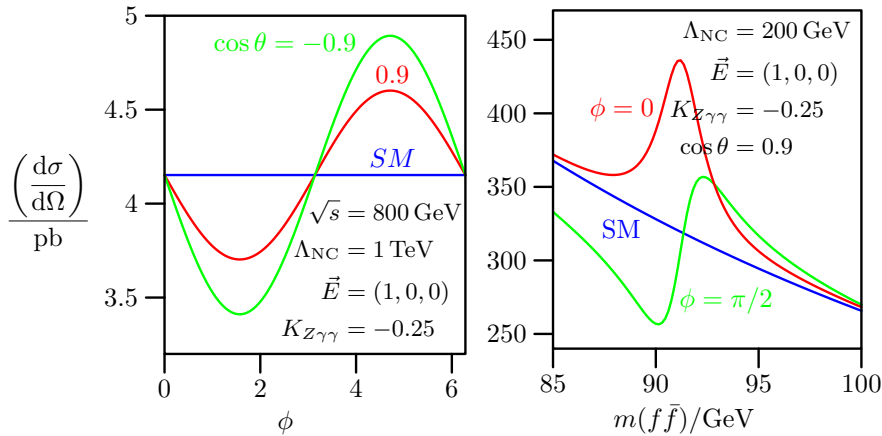

  \begin{center}
  \includegraphics{ncpc_phi}
  \includegraphics{ncpc_res}
  \end{center}
  \caption{\label{fig:zres}%
    \emph{Left:} Azimuthal dependence of the amplitudes. The
    horizontal line is the SM differential cross section.
    \emph{Right:} The $Z$ resonance in the interference of SM and
    $\mathcal{O}(\theta)$ NCSM amplitude plotted for different values of 
    the azimuth angle.}
\end{figure}

\begin{figure}
  \begin{center}
  \includegraphics{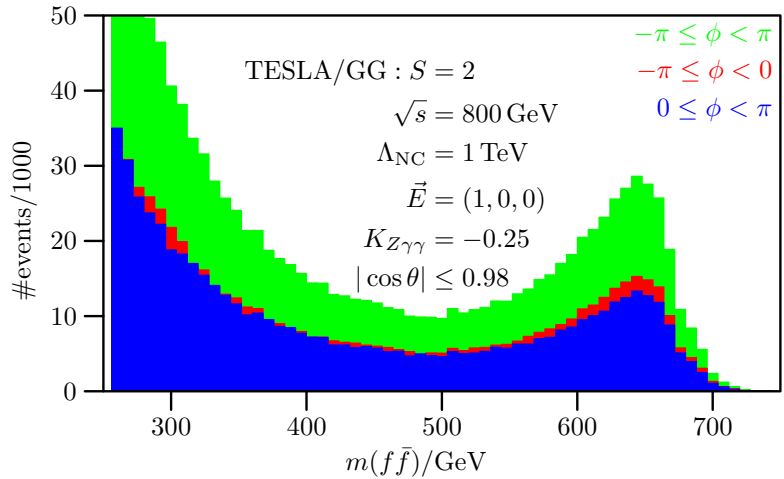}
  \end{center}
  \caption{\label{fig:s800fl}%
    Number of events per year in the two halfspheres $\phi \gtrless 0$
    for $\sqrt{s}=800\,\GeV$ in the $S=2$ mode. A cut has been made
    $0.02$ radians around the beam axis.}
\end{figure}

Due to helicity conservation in the SM, the only nonvanishing
combinations for the processes $\gamma\gamma\to f\bar f$ with massless
fermions are~$(\pm,\mp) \to (\pm,\mp)$ and~$(\pm,\mp) \to (\mp,\pm)$.
Thus a deviation from the SM can only be seen, if the two photon
helicities add up to a total of two. Therefore the photon collider has
to be run in the $d$-wave mode ($S=2$), while the preferred mode for
Higgs physics is the $s$-wave ($S=0$).  Fortunately, one can easily
switch between them. For both SM diagrams, there are two $O(\theta)$
contributions, where one SM vertex is replaced by the corresponding
NCSM vertex, while there is a new contact term as well as two new
$s$-channel diagrams containing the TGCs. A nonvanishing interference
appears only for non-zero $E^1$ or $E^2$. The explicit expressions for
the different helicity amplitudes are published elsewhere~\cite{thojr}.


As cross checks, we verified numerically the amplitudes calculated
within two completely different formalisms -- helicity amplitudes
vs. Dirac spinors using the optimizing matrix element compiler
\texttt{O'Mega/WHIZARD}~\cite{Omega} -- to agree within numerical
accuracy. 

In the NCSM, the $Z$ can be produced resonantly in photon collisions.
The right figure \ref{fig:zres} 
shows the resonance structure around the $Z$ pole. Unfortunately this
can not be well observed at the photon collider because of poor
statistics in the photon spectrum for low energies. 
In figure~\ref{fig:zres} on the left, the differential cross section at 
$\sqrt{s} = 800\,\GeV$ is shown plotted against the
azimuth angle $\phi$. From this, one sees that the integrated cross section
is the same for the pure SM and the interference, as the interference
goes like $\propto \sin(\phi + \phi_0)$, where $\phi_0$
is a phase which depends on the spatial orientation of the
noncommutativity. However, in the polarized cross section it is
possible to scan over the final state fermions to look for angular
dependent deviations from the SM prediction. 

To get a realistic result, one has to fold the cross
section with the photon spectrum produced in Compton scattering for
the photon collider, for which the program {\tt Circe 2.0} has
been used \cite{Circe}. In figure \ref{fig:s800fl} we show the number of
binned events over the invariant mass of the fermion pair, assuming an 
integrated luminosity of ${\mathcal{L}} = 2000 \;\fb^{-1}$. The cross
section and angular variation away from the $Z$ resonance do not
depend very much on the specific value of the TGCs. A signal
could always be seen if the scale $\Lambda_{\text{NC}}$ is even a
little bit higher than the CMS energy of the linear collider.
It is not consistent to consider the amplitudes and cross section
for collider energies higher than the scale $\Lambda_{\text{NC}}$,
since higher orders in $\theta$ can only be neglected if
$s/(\Lambda_{\text{NC}})^2 \lesssim 1$. 


In conclusion, the NCSM offers a variety of new phenomena due to the
presence of an antisymmetric rank-2 spurion field which breaks
Lorentz invariance at a scale $\Lambda_{\text{NC}}$. This violation of
angular momentum conservation is a generic signal for
noncommutativity, which leads to deviations from the isotropic
distribution of final state particles around the beam axis. We have focused
on fermion pair production at a future photon collider. Polarization
is shown to be a mandatory ingredient to be sensitive for signals of
noncommutative theories.  In a conservative
estimate, a photon collider will be sensitive to scales of the order of 
$\Lambda_{\text{NC}} \sim 1\,\TeV$, but if enough data will be
available, the sensitivity may even be better by a factor three to
five.     

\begin{figure}
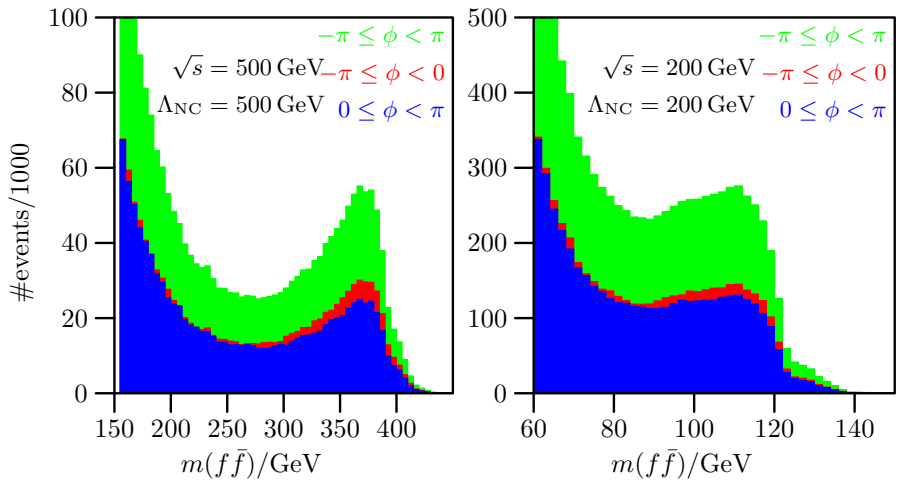

  \begin{center}
  \includegraphics{ncpc_500}%
  \includegraphics{ncpc_200}
  \end{center}
  \caption{\label{fig:s500fl/s200fl}%
    Number of events per year in the two halfspheres $\phi \gtrless 0$
    for $\sqrt{s}=500\,\GeV$ and $\sqrt{s}=200\,\GeV$ in the $S=2$ mode.}
\end{figure}


\vspace{-1.5mm}



\section*{References}

\baselineskip13pt
\begin{thebibliography}{99}

\bibitem{SeiWit}
  N.~Seiberg, E.~Witten, 
  JHEP 9909 (1999) 032
  [hep-th/9908142]

\bibitem{Wess:pr}
  J.~Wess,
  Commun.{} Math.{} Phys.{}  \textbf{219} (2001) 247.

\bibitem{NCSM}
  X.~Calmet et al.,
  Eur.\ Phys.\ J.\ {\bf C23} (2002) 363
  [hep-ph/0111115].

\bibitem{TGC}
  G.~Duplan{\v c}i\'c, P.~Schupp, and J.~Trampeti\'c, 
  Eur.\ Phys.\ J.\ {\bf C32} (2003) 141
  [hep-ph/0309138] and references therein.

\bibitem{thojr}
  T.~Ohl, J.~Reuter,
  Phys.{} Rev.{} \textbf{D} (in print)
  [hep-ph/0406098].

\bibitem{Omega}
  T. Ohl, hep-ph/0011243;
  M.~Moretti, T.~Ohl, J.~Reuter,
  hep-ph/0102195;
  \texttt{http://whizard.event-generator.org};   W.~Kilian, T.~Ohl, 
  J.~Reuter,
  hep-ph/0708.4233.

\nopagebreak

\bibitem{Circe}
  T.~Ohl, 
  Comput.\ Phys.\ Commun.\ {\bf 101} (1997) 269
  [hep-ph/9607454];
  T.~Ohl, {\tt Circe 2.0}, WUE-ITP-2002-006.



\end{thebibliography}
\end{document}